\documentclass[aps,prd,preprint,nofootinbib]{revtex4-1}
\usepackage{amsmath,amssymb,graphics,graphicx,color}
\usepackage{scrextend}
\usepackage{hyperref}
\usepackage{float}
\usepackage{xcolor}
\usepackage[utf8]{inputenc}
\usepackage[english]{babel}
\usepackage{slashed}

\begin{document}

\title{Constraining the Active-to-Heavy-Neutrino transitional magnetic moments associated with the $Z'$ interactions at FASER$\nu$.}
\author{Kingman Cheung$^{a,b,c}$, C.J. Ouseph$^{a,b}$}
\affiliation{
  $^a$ Department of Physics, National Tsing Hua University, Hsinchu 30013,
  Taiwan\\
  $^b$ Center for Theory and Computation, National Tsing Hua University, Hsinchu 30013, Taiwan \\
  $^c$ Division of Quantum Phases and Devices, School of Physics,
  Konkuk University, Seoul 143-701, Republic of Korea
}
\date{\today}

\begin{abstract}
We investigate the effects of the transitional magnetic dipole moment of the active-to-heavy-neutrino associated with a new neutral gauge boson $Z'$  on neutrino-nucleon scattering at the ForwArd Search ExpeRiment-$\nu$ (FASER$\nu$). 
We consider the neutral-current 
neutrino-nucleon scattering ($\nu A\rightarrow N A$) as the production mechanism of the heavy neutrino $N$ at FASER$\nu$ and estimate the sensitivity reach on the magnetic moment coupling $\mu_{\nu_\alpha}$ for a range of heavy neutrino mass 
($1\,{\rm GeV} < M_{N} < 70 {\rm GeV}$). In this study, we consider three benchmark models, in which the heavy neutrino is coupled to $L_e$, $L_\mu$
or $L_\tau$ doublet, respectively.
\end{abstract}

\maketitle

\section{Introduction}
Heavy neutrinos beyond the current known species of active neutrinos
are predicted in a number of grand unified theories (GUT), such as GUT
based on $SO(10)$, and are motivated as the natural explanation why
the active neutrinos are so light compared with other matter fermions.
One of the natural models is the seesaw model \cite{Yanagida:1979as,Mohapatra:1979ia}, 
in which the heavy neutrinos
are right-handed neutrinos of mass $O(10^{11-14})$ GeV, that explains
the order of active mass to be $y (vev)^2/M_N$, where the $vev \sim 10^2$
GeV is the electroweak vacuum expectation value. There are other
extensions \cite{Dev:2009aw,Deppisch:2004fa} of the ideas of the
seesaw mechanism that predict intermediate mass
neutrinos and can also be as light as a few GeV's. Here in this work we 
focus on the heavy neutrinos of mass in the range of $1-70$ GeV.

FASER \cite{FASER:2019dxq,Kling:2021gos,FASER:2019aik,FASER:2021ljd,FASER:2021cpr,FASER:2020gpr,FASER:2018bac,Feng:2022inv} is one of the neutrino-nucleon scattering experiments being approved and coming up very soon
in the next run of the LHC. It is located approximately 480 meters 
down the beam direction from the ATLAS detector interaction point (IP). 
It is well known that a large number of hadrons are produced along the 
beam direction, such as pions, kaons, and other hadrons. 
During the flight, these hadrons will decay thus producing neutrinos 
of all three flavors at very high energies up to a few TeV. 
The neutrinos so produced can pass through hundred meters of rock and concrete between the ATLAS IP and the FASER detector, whereas most other particles will be either deflected or absorbed before reaching FASER. 
FASER is unique in terms of the energy range of the neutrinos 
as beams participating in $\nu-A$ scattering. It ranges from a hundred GeV to a few TeV with the energy spectrum peaked at a few hundreds GeV. 
Other similar setups, such as DUNE near detector, have lower 
energy ranges. It is an important avenue of the energy frontier for neutrino physics, which is an area often considered to be on the intensity frontier. 

The FASER detector will be complemented with a new component called FASER$\nu$\cite{FASER:2019dxq}. It is a 25cm × 25cm × 1.35m emulsion detector, consisting of 1000 layers of emulsion films interleaved with 1-mm-thick tungsten plates with mass 1.2 tons. The main goal of FASER$\nu$ is to distinguish various flavors of neutrinos. The FASER$\nu$ has the ability to detect both charged-current(CC) \cite{FASER:2019dxq} and neutral-current (NC) interactions \cite{Ismail:2020yqc,Ansarifard:2021elw}. The total cross-section $\sigma_{\nu A}$ (where $A$ denote the target nucleus) of the neutrino scattering at the FASER$\nu$ detector can be expressed as $\sigma_{\nu A} = p\sigma_{\nu p}+n\sigma_{\nu n}$, where $\sigma_{\nu p}$ and $\sigma_{\nu n}$ are the neutrino-proton and neutrino-neutron scattering cross sections, and $p$ and $n$ are the number of protons and neutrons in the tungsten atom, respectively. Various new physics scenarios at FASER and FASER$\nu$ detector were reported in literature\cite{Jodlowski:2020vhr,Feng:2022inv,Feng:2017vli,Kling:2018wct,Feng:2018pew,Deppisch:2019kvs,Cheung:2021tmx,Bakhti:2020szu,Jho:2020jfz,Okada:2020cue,Bahraminasr:2020ssz,Kelly:2020pcy,Falkowski:2021bkq,FASER:2018eoc,Cottin:2021lzz,Ismail:2021dyp}.

In this work, we consider a scenario of a heavy neutrino $N$ with 
the presence of a light $U(1)$ gauge boson $Z'$\cite{Das:2022rbl,Das:2019fee,Chiang:2019ajm}.
The heavy neutrino $N$ can couple to the active neutrinos via a
transitional magnetic dipole moment type interaction of the 
$Z'$ boson:
$\overline{N} \sigma^{\mu\nu} \nu_\alpha\, Z'_{\mu\nu}$, where $Z'_{\mu\nu}$ 
is the field strength of the $Z'$ boson. Such an interaction enables
sizable production of the heavy neutrinos at FASER$\nu$ and also 
detectable decays via $N \to \nu_\alpha Z' $ followed by 
$Z'$ decays into standard model (SM) fermions. In literature, the works that only
considered the $N$ decays via the SM $W$ and $Z$ often obtained very long
decays certainly outside the detectors. The advantage of including 
the $Z'$ interaction with the heavy neutrinos is to enable either
prompt or detectable long decays within the FASER detector.
Here the $Z'$ boson also couples to the SM fermions $q, \nu, l$. 
For completeness we also include the couplings of $N$ to the SM 
charged leptons and neutrinos via the $W$ and $Z$ bosons, respectively, 
although their contributions to production and decays of $N$ 
are very small.

We identify the events of heavy neutrinos through its decays 
into a pair of charged leptons, either in 
prompt or displaced decays, as well as into hadrons. 
The smaller the coupling $\mu_{\nu_\alpha}$
the longer the decay length will be.  Thus, using the 
length of the whole FASER detector we can estimate the
range of $\mu_{\nu_\alpha}$ that FASER and FASER$\nu$ can probe. 
The final result would be the sensitivity that FASER$\nu$ can
reach in this scenario.

Note that another scenario of sterile neutrino was considered in Ref.~\cite{Jodlowski:2020vhr}, in which the
sterile neutrino is produced through the magnetic-dipole operator 
via the active neutrino up-scattering, mostly via 
the electron scattering channel $\nu e^- \to N e^-$. It was 
shown that \cite{Jodlowski:2020vhr} the High-Luminosity LHC 
can search for sterile neutrinos with an upgraded FASER$\nu$ experiment.

The organization of the work is arranged as follows. We describe the 
effective interactions considered in the next section. In Sec.~III, 
we calculate the production rates of the heavy neutrino $N$ via
$\nu  A$ scattering at FASER$\nu$. In Sec.~IV, we calculate the number of
events for the decays of the heavy neutrino and then estimate the 
sensitivity reach at FASER$\nu$. We conclude in Sec. V.

\section{The Heavy Neutrino Model}
The new matter content includes a heavy neutrino $N$ and a
$U(1)$ gauge boson $Z'$.
The effective Lagrangian describing the interactions of $N$ and $Z'$ 
with the standard model (SM) gauge bosons \cite{Drewes:2015iva,Shrock:1978ft,FMMF:1994yvb,Pascoli:2018heg,Alva:2014gxa,Atre:2009rg} and other particles  \cite{He:1991qd,Hewett:1988xc,Leike:1998wr,Langacker:2008yv,Lee:2008zzl}can be written as
\begin{eqnarray}\label{Eq.1}
\mathcal{L}_{\rm eff} &=& \sum_{\alpha=e,\mu,\tau} \Biggr [ \mu_{\nu_\alpha} \overline{N}\sigma^{\mu\nu}\nu_{\rm \alpha} Z'_{\mu\nu} 
 - \frac{g}{\sqrt{2}}V_{\alpha N}\overline{N} \gamma^\mu P_L l_{\alpha} W_\mu^+  - \frac{g}{ \cos\theta_w} V_{\alpha N}\overline{N} \gamma^\mu P_L\nu_{\alpha} Z_\mu +\rm H.c. \Biggr ] \nonumber \\
 & -& \sum_{q,\nu,l} \left[ g_{\rm q}\bar{q}\gamma^{\mu}q+ g_{\nu} \bar{\nu} \gamma^{\mu} P_L \nu  +g_{l} \bar{l} \gamma^{\mu} l \right ] Z'_{\mu} \;,
\end{eqnarray}
where $g=e/\sin\theta_w$, $\theta_w$ is the Weinberg angle, 
$Z'_{\mu\nu}$ is the field strength of the $Z'$ boson, 
the sum over $q=u,d$,
$\nu=\nu_e,\nu_\mu, \nu_\tau$ and $l =e,\mu,\tau$, and the mixing parameter
$V_{\alpha N}$ denotes the mixing between $\nu_\alpha$ and $N$. 
Here $\mu_{\nu_\alpha}$ denotes the coupling of the active-to-heavy-neutrino transitional magnetic dipole moment, for which the first term in the Lagrangian Eq.~(\ref{Eq.1}) is valid up to a cut-off energy scale $\Lambda$. 
An interpretation of $\mu_{\nu_\alpha}$ above the electroweak scale requires
a Higgs insertion so that the transitional magnetic dipole interaction
described in Eq.~(\ref{Eq.1}) originates from a dimension-6 operator.
The $\mu_{\nu_\alpha}$'s are therefore envisioned to be of the order of 
$1/\Lambda^2$ in form of $\mu_{\nu_\alpha} \sim \frac{e.vev}{\Lambda^2}$.

The effective Lagrangian consists of two widely used beyond the SM models: 
(i) the renormalizable $Z'$ model with flavor-conserving quark, lepton and neutrino interactions \cite{He:1991qd,Hewett:1988xc,Leike:1998wr,Langacker:2008yv,Lee:2008zzl} 
and (ii) an effective/simplified extension by introducing a right-handed (RH) neutrino, which is a singlet under the SM gauge symmetry\cite{Pascoli:2018heg,Alva:2014gxa,Atre:2009rg}, with 
a transitional dipole moment connecting the active and the heavy neutrino.
Without loss of generality, we assume that $q = u, d$ have equal coupling strength $g_q$ with the $Z'$, $\nu = \nu_e,\nu_\mu, \nu_\tau$ have equal strength $g_\nu$ with the $Z'$, and equal coupling strength $g_l$ for 
$l=e,\mu,\tau$.

 \begin{figure}[t!]
	\centering
	\includegraphics[width=10cm,height=7cm]{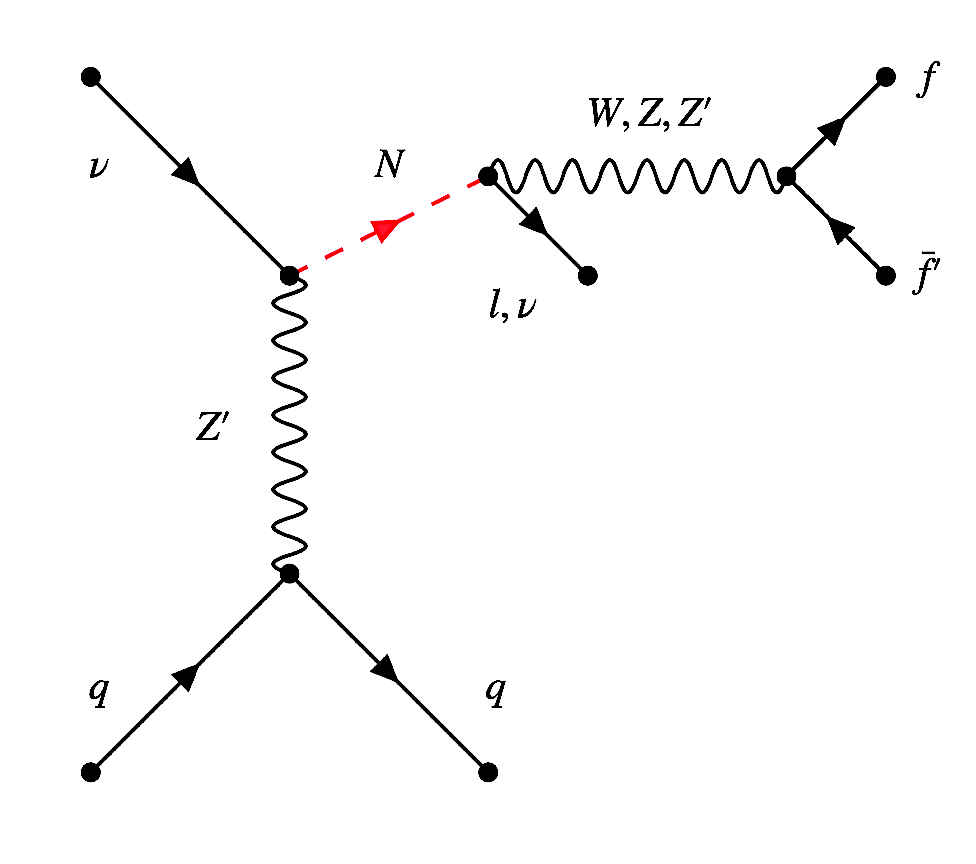}
	\caption{ \small \label{fig1}
	Production and decays of the heavy neutrino $N$ at FASER$\nu$. ($\nu q \rightarrow N q$ followed by $N \rightarrow \nu f \bar{f}'$, where
	$f,f'$ are SM fermions.)}
\end{figure}

\section{Heavy Neutrino production at FASER$\nu$}

\begin{figure}[th!]
	\centering
	\includegraphics[width=16cm,height=12cm]{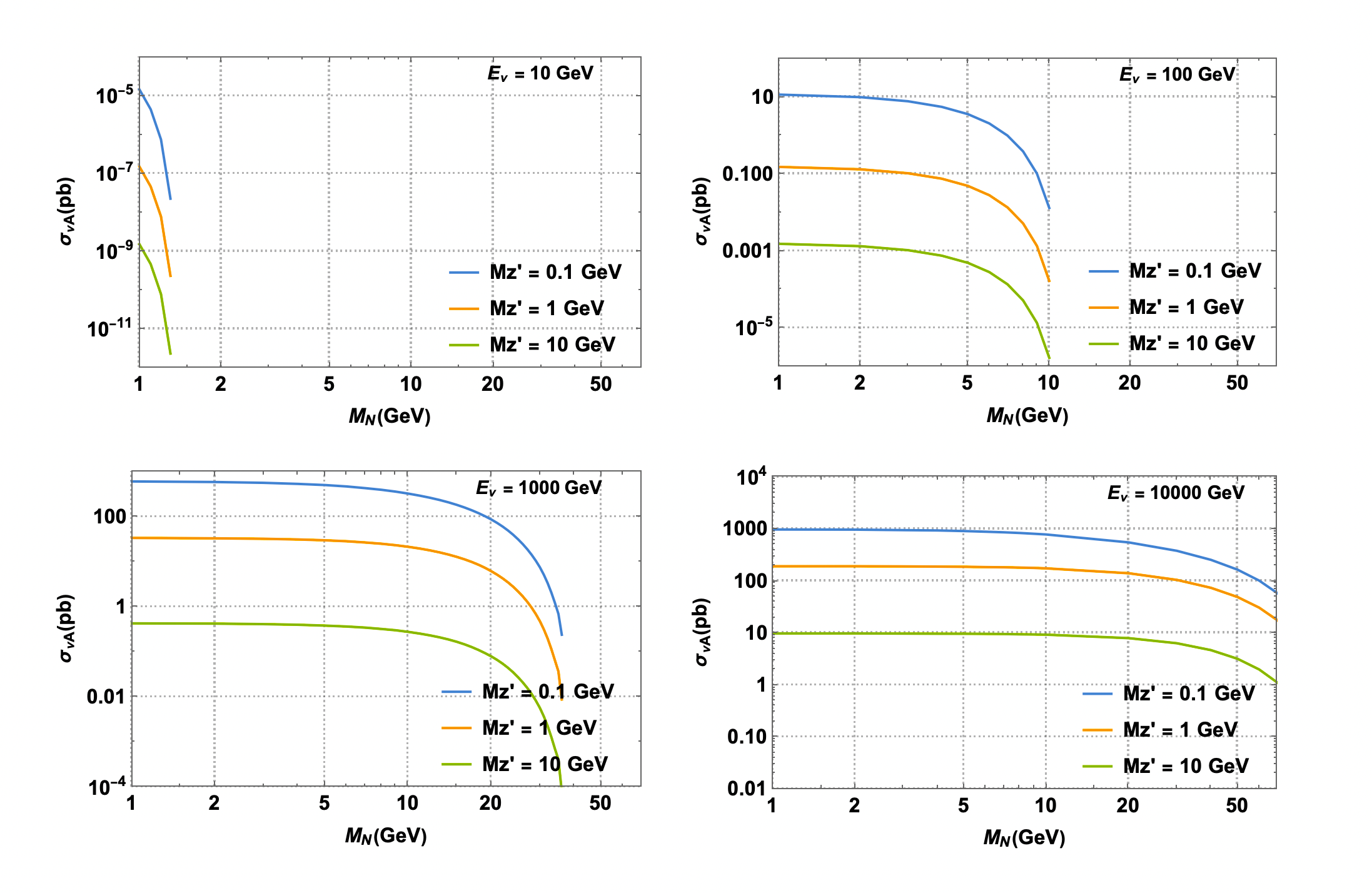}
	\caption{ \small \label{fig2}
	Production cross sections of the heavy neutrino $N$ at FASER$\nu$: $\sigma_{\nu A} = p\sigma_{\nu p}+n\sigma_{\nu n}$, computed
	with $\Lambda = 1000~\rm GeV$, and $g_q = 1$. 
	Here $A$ is the tungsten nucleus with $p$ protons and $n$ neutrons. 
	The energy $E_\nu$ of the incoming neutrino beam is $E_\nu$ = 10 GeV
	(top left),  $E_\nu$ = 100 GeV (top-right), $E_\nu$ = 1000 GeV (bottom-left), and $E_\nu$ = 10000 GeV. (bottom-right), respectively. }
\end{figure}

In this section, we discuss heavy neutrino production at FASER$\nu$ detector from the neutrino-nucleon scattering $ \nu A\rightarrow N A $ mediated by a new neutral gauge boson $Z'$. The heavy neutrino is produced at FASER$\nu$ and it will decay into SM particles 
($ N \rightarrow \nu l^+ l^- $, $ N \rightarrow \nu q \bar q $, $ N \rightarrow \nu \nu \bar \nu $). As shown in Eq.~(\ref{Eq.1}) the 
heavy neutrino $N$ has a dipole-like vertex with the active neutrino and the new gauge $Z'$ boson, which resembles the transitional magnetic moment. The Feynman diagram of the production and decay process of heavy neutrino $N$ is shown in Fig.~\ref{fig1}. Discovery potential for heavy neutral leptons at FASER through the SM gauge boson decays was discussed in \cite{Jodlowski:2020vhr,FASER:2018eoc,Kling:2018wct}.

We use $\rm MadGraph5aMC@NLO$ \cite{Alwall:2011uj,Alwall:2014hca} for the computation of fixed-target neutrino-nucleon scattering $ \nu A\rightarrow N A $ cross-sections. We build the model file for Eq.~(\ref{Eq.1}) using FeynRules \cite{Alloul:2013bka}. 
For the production process we consider 3 different neutral gauge boson masses, $M_{Z'}=0.1,1,10~\rm GeV$, and the heavy neutrino mass 
from 1 to 70 GeV ($M_N<  M_{w}$). 
FASER$\nu$ is providing a unique region of neutrino energy 
1 GeV -- 10 TeV for heavy neutrino production. Note that we fixed the value of tree level $Z'$-quark coupling 
strength $g_q=1$ and the cut-off energy scale value $\Lambda=1000$ GeV. 
The production cross section simply scales with 
$g_q^2$ and $1/\Lambda^4$.
Figure~\ref{fig2} shows the neutrino-nucleon fixed target cross-sections 
for different neutrino beam energies $E_\nu = 10 - 10000$ GeV. In each case, the cross-section decreases with increment of both $M_{Z'}$ and $M_N$.
Here we have assumed the benchmark detector specification 
made of tungsten with 
dimensions $25\,{\rm cm} \times 25\, {\rm cm}\times 1\, {\rm m}$ at the 
14 TeV LHC with an integrated luminosity of $L = 150\,{\rm fb}^{-1}$. 
We use the neutrino fluxes and energy spectra obtained in \cite{FASER:2019dxq,Kling:2021gos}
to study the neutrinos that pass through FASER$\nu$. We find that 
the muon neutrinos are mostly produced from charged-pion decays, electron neutrinos from hyperon, kaon and $D$-meson decays, and tau neutrinos from $D_s$ meson decays. With average energies ranging from 600 GeV to 1 TeV, the spectra of the three neutrino flavors cover a broad energy range.

\begin{figure}[h!]
	\centering
	\includegraphics[width=16cm,height=12cm]{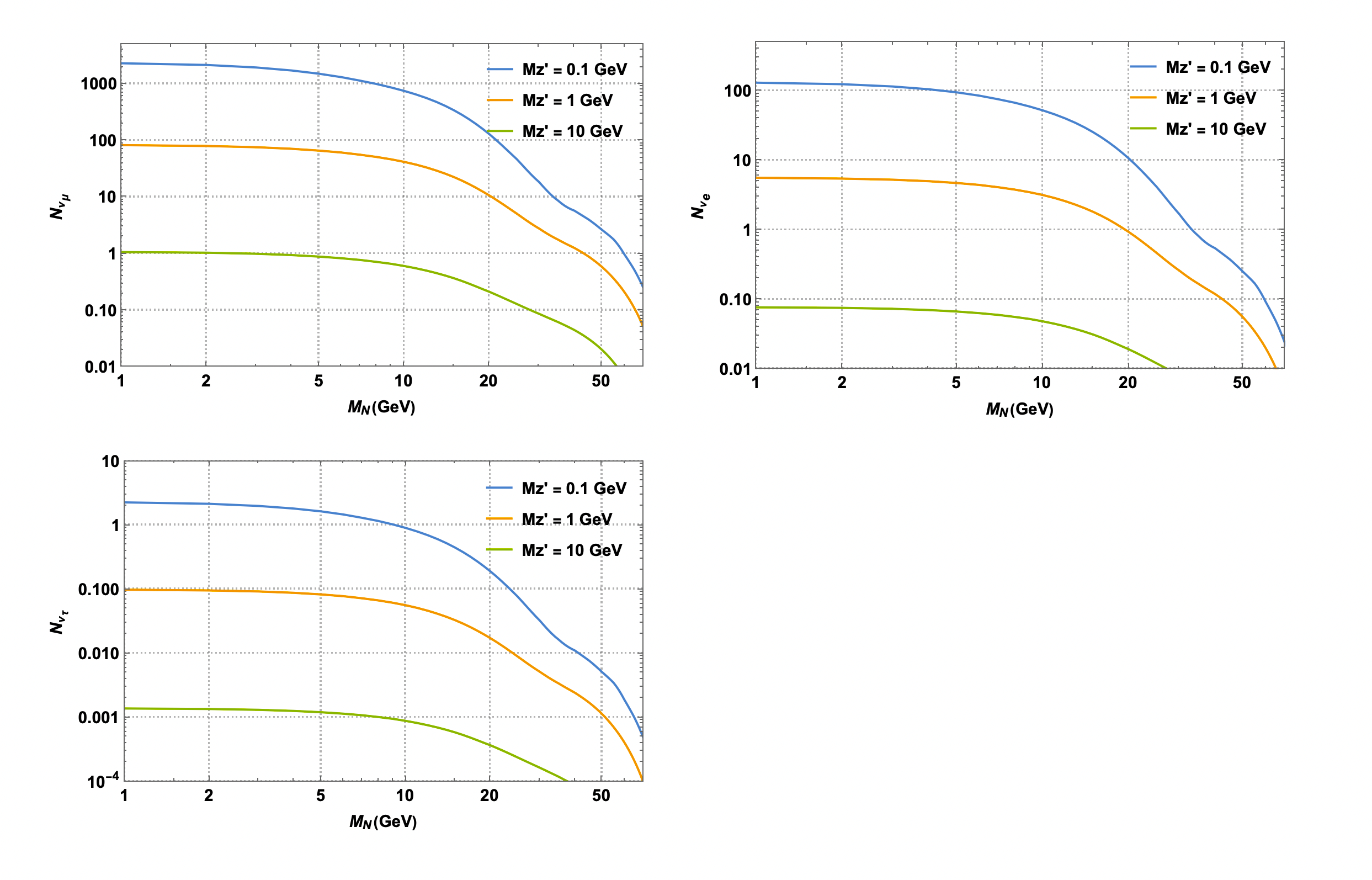}
	\caption{ \small \label{fig3}
	Total number of heavy neutrinos produced by scattering with the tungsten 
	target at FASER$\nu$ with $\nu_{\mu} A \rightarrow N A$ scattering
	(top-left), with the $\nu_{e} A \rightarrow N A$ scattering (top-right), and with $\nu_{\tau} A \rightarrow N A$ scattering (bottom-left).} 
\end{figure}

The expected number of heavy neutrino events from the three flavors of active neutrinos with three different $Z'$ masses in 
FASER$\nu$ is shown in 
Fig.~\ref{fig3}. The highest number of heavy neutrino events come in 
 with the muon neutrino $\nu_\mu$ scattering with tungsten nucleus($\nu_{\mu} A \rightarrow N A$), while the lowest one is from the tau neutrino $\nu_\tau$ scattering ($\nu_{\tau} A \rightarrow N A$), and those from  
 the electron neutrino $\nu_e$  scattering is in between.
 Total number of events follow the same behavior as the scattering cross-section. 
 We can summarize Fig.~\ref{fig3} as follows, $N_{\nu_\tau}(M_{Z'},M_{N},\mu_{\nu_\tau},g_{q}) < N_{\nu_e}(M_{Z'},M_{N},\mu_{\nu_e},g_{q}) < N_{\nu_\mu}(M_{Z'},M_{N},\mu_{\nu_\mu},g_{q})$.

\begin{figure}[th!]
	\centering
	\includegraphics[width=16cm,height=8cm]{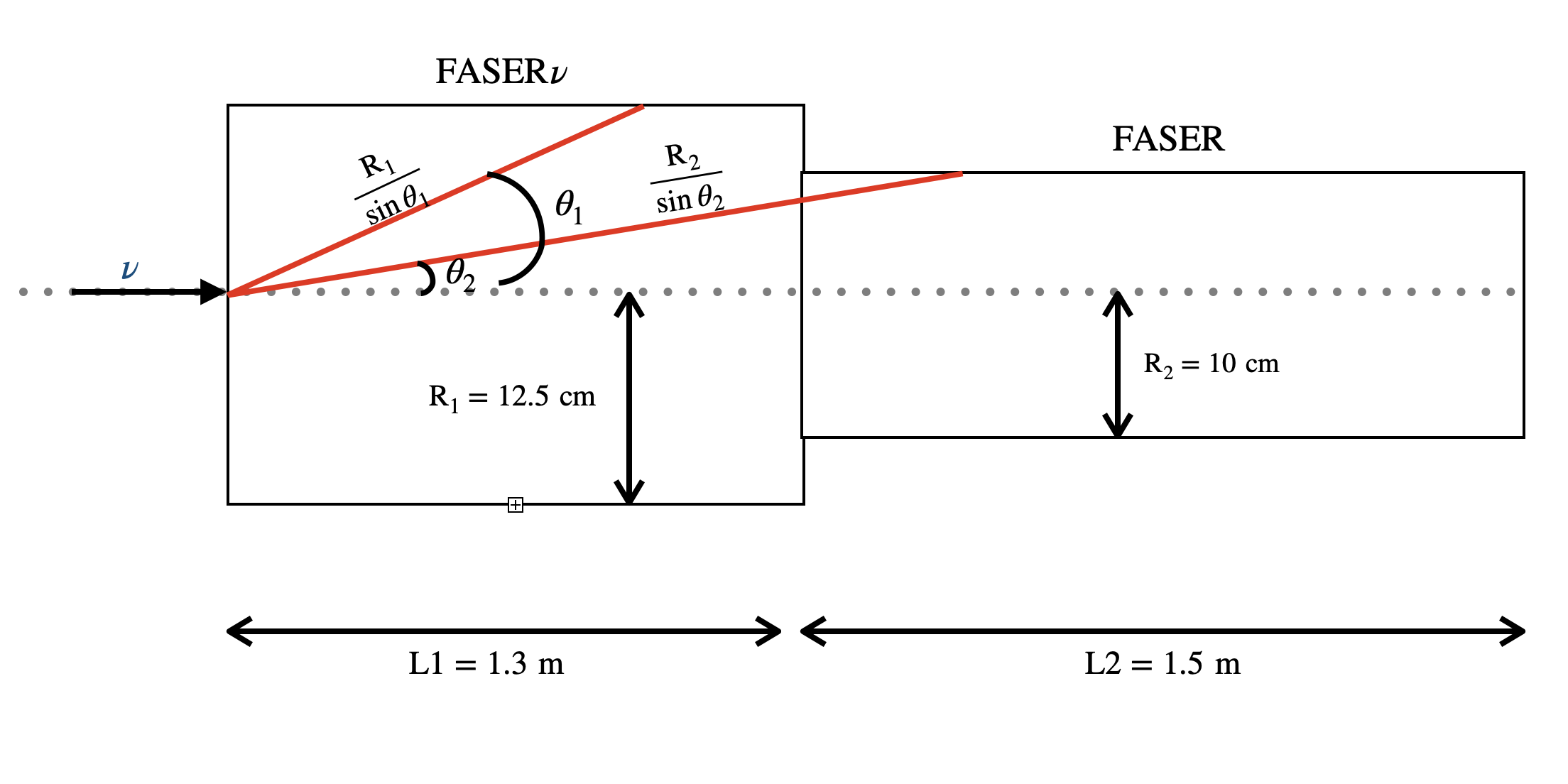}
	\caption{ \small \label{fig4}
	Schematic diagram of FASER$\nu$-FASER~\cite{Feng:2017vli,Kling:2018wct,Feng:2018pew,Deppisch:2019kvs,Cheung:2021tmx,Okada:2020cue,Bahraminasr:2020ssz,Kelly:2020pcy,Falkowski:2021bkq,FASER:2018eoc} }
\end{figure}

\section{FASER$\nu$-FASER Sensitivity towards active to heavy Neutrino transitional magnetic moments }
The heavy neutrino produced at the FASER$\nu$ detector could decay into 
visible channels -- charged leptons and quarks. The probability of heavy neutrino decay within the detector volume is given by\cite{Kling:2018wct},
\begin{equation}\label{Eq.2}
    \mathcal{P}_{\rm detc}=1-\exp(\frac{-d}{\beta c\tau}) \;,
\end{equation}
where $d$ is defined as $d \equiv {\rm min}
(\Delta, \frac{R}{\sin\theta})$, $\Delta$ is the length of the detector, $R$ is the radius of the detector and $\theta$ is the angle between the outgoing heavy neutrino $N$ along with horizontal beam axis. A schematic diagram of the detector is given in Fig.~\ref{fig4}.
The decays of the heavy neutrino $N$ into SM particles are 
calculated using the $\rm MadGraph5aMC@NLO$ \cite{Alwall:2011uj,Alwall:2014hca,Alwall:2014bza}.
The angular distribution of the heavy neutrino produced at FASER$\nu$ 
is extracted from the output files with the extension $\rm ``.lhe"$.

We generate 10,000 events using $\rm MadGraph5aMC@NLO$ for the computation of the angular distribution of the heavy neutrino with respect to the horizontal axis. We consider different sets of the heavy neutrino mass ranging from 1 GeV to 70 GeV and the $Z'$ mass ($M_{Z'}$=0.1, 1, 10 GeV) with neutrino energy $E_\nu$ ranging from 10 GeV to 10 TeV. We determine 
the fraction of events at a particular scattering angle from the angular distribution of heavy neutrino $N$. 
\begin{figure}[th!]
	\centering	\includegraphics[width=16cm,height=10cm]{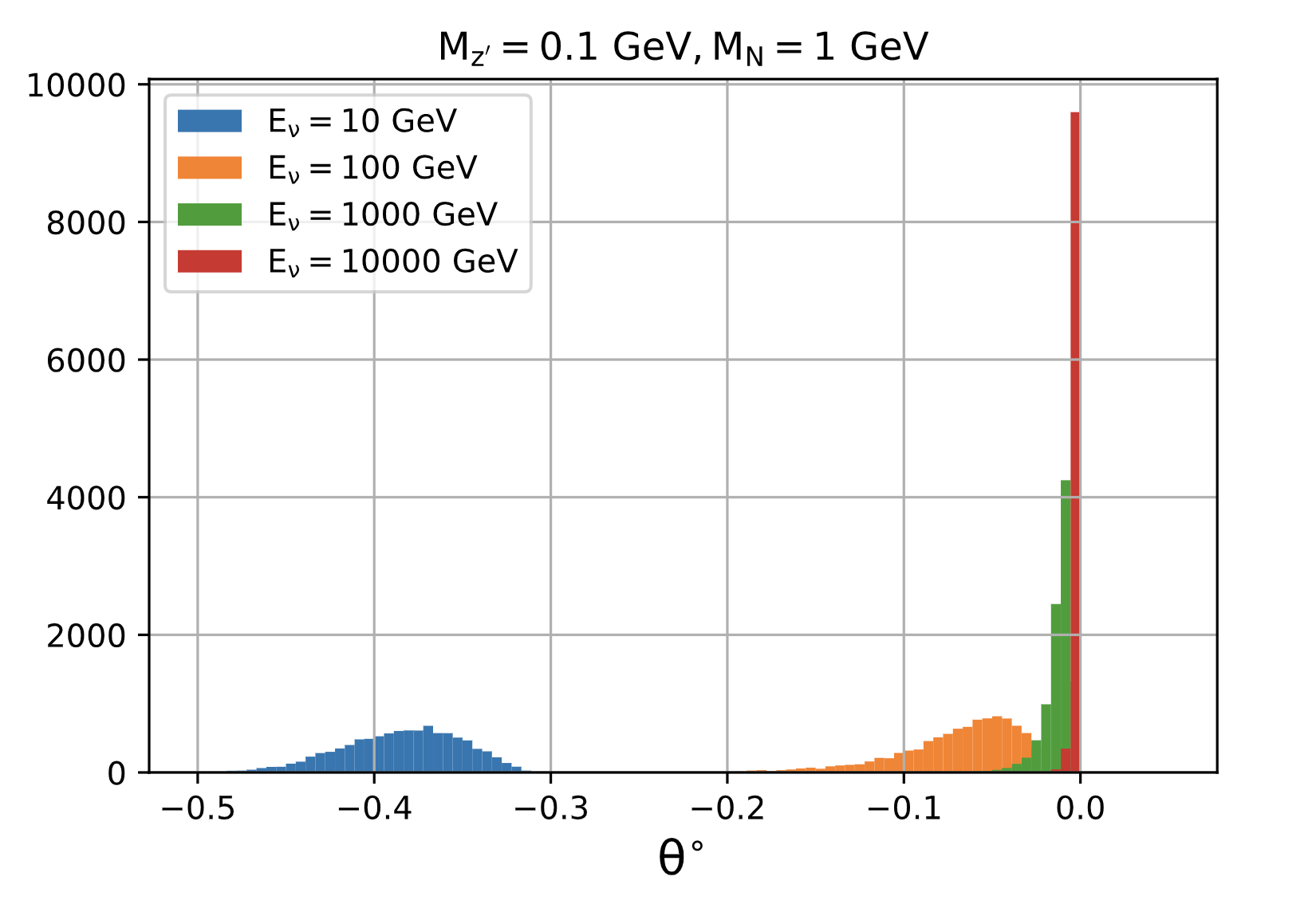}
	\caption{ \small \label{fig5}
Angular distribution of the heavy neutrino of mass $M_{N}$ = 1 GeV with different sets of neutrino beam energies. The mediator $Z'$ mass is set 
at $M_{Z'}$ = 0.1 GeV 	}
\end{figure}
Figure~\ref{fig5} shows the angular distributions of the 
heavy neutrino versus the scattering angle $\theta$ for 
$E_\nu = 10 - 10000$ GeV.
As expected the high energy neutrino so produced at the 
FASER$\nu$ travels very close to the beam axis. 
The higher the incoming neutrino beam energy
and the heavier the $N$, the closer to the beam axis the heavy 
neutrino $N$ travels (i.e. the smaller the scattering angle). 
The variation with different $M_{Z'}$ is not significant.
More details about the angular distribution are given in appendix \ref{app}.

The total number of the heavy neutrino decays detected within the  FASER$\nu$ detector is given by
\begin{equation}\label{Eq.3}
N^{\rm detc}_{\alpha} = N_{\alpha}^{\rm Prod}(\nu_{\alpha} 
A \rightarrow N A)\times  \mathcal{P}_{detc}\times {\rm BR}(N\rightarrow \nu_{\alpha}~X~X) \;,
\end{equation}
where $N_{\alpha}^{\rm Prod}(\nu_{\alpha} A \rightarrow N A) $ denotes the the total number of heavy neutrino events produced at FASER$\nu$ from the $\nu_\alpha$ as shown in Fig~\ref{fig3}, $\rm \mathcal{P}_{detc} $ is the probability of heavy neutrino decay within the detector volume Eq.(\ref{Eq.2}), and ${\rm BR}(N\rightarrow \nu_{\alpha}~X~X) $ represents the branching ratio of heavy neutrino $N$ into $\nu_\alpha X X $.
There are two possible detecting channels in the $N$ decays: (i) $N \rightarrow \nu l^+ l^-$ (a pair of charged leptons), and  (ii) $N \rightarrow \nu q \bar q$ (a pair of quarks appearing as hadrons).

The final-state charged leptons and quarks from heavy neutrino $N$ decays are detectable in the $\rm FASER\nu$ and FASER detector. We also assume that backgrounds can be reduced to negligible levels for 
both cases of prompt and displaced decays of the heavy neutrino.
Thus, we present the sensitivity results in terms of 3-signal-event 
contour curves which correspond to 95\% C.L. limits with zero background \cite{Tian:2022rsi}. We consider 3 benchmark models for the evaluation of the FASER and FASER$\nu$ sensitivity for the active-to-heavy-neutrino transitional magnetic moment. The three benchmark models are discussed below.

\subsection{Benchmark Model I (BM-I): 
Heavy Neutrino $ N$ and $Z'$ mixed with $\mu$ and $\nu_\mu$}

\begin{figure}[th!]
	\centering
	\includegraphics[width=16cm,height=14cm]{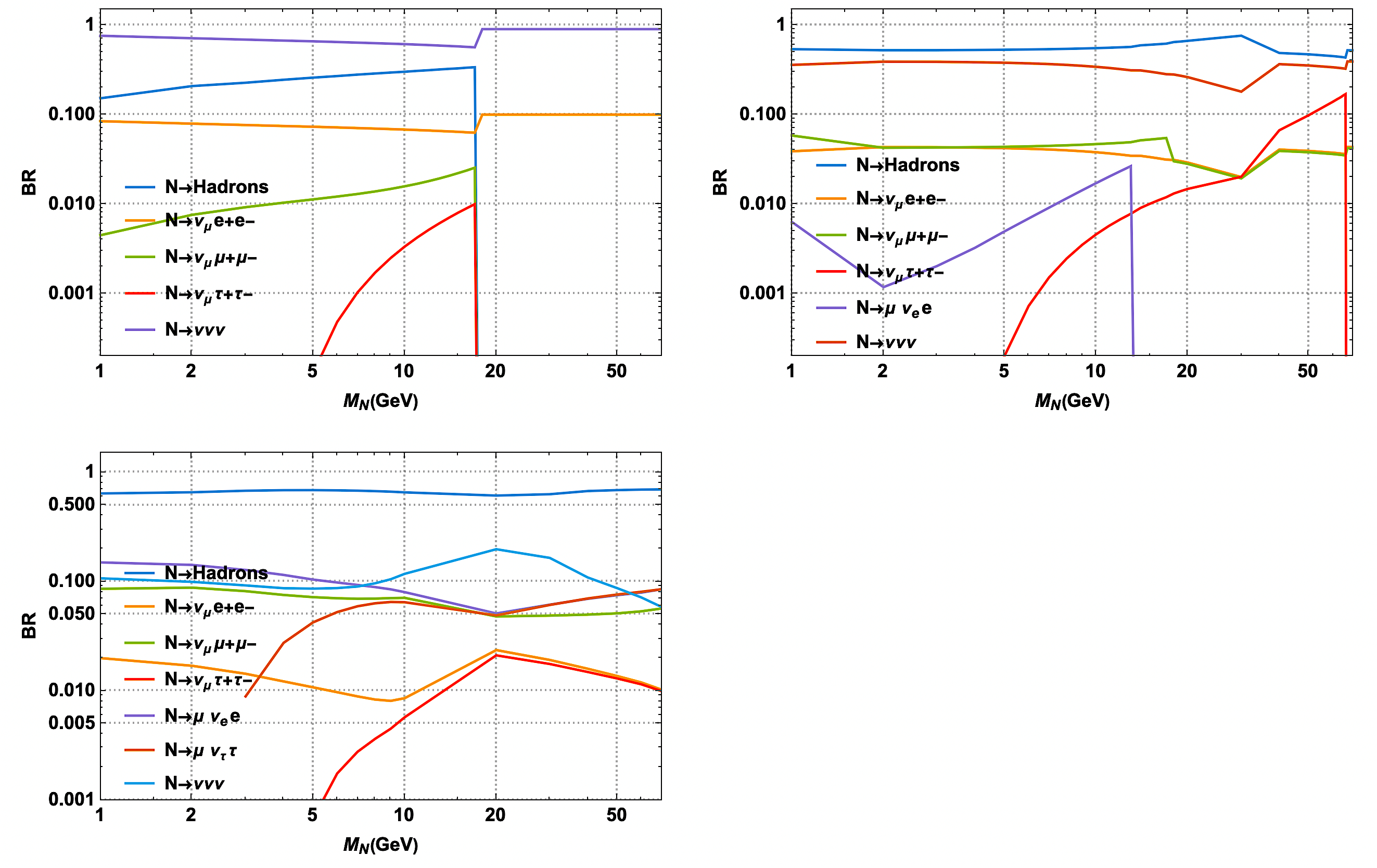}
	\caption{ \small \label{fig6}
    Dominant branching fractions of the heavy neutrino $N$ as a function of its mass $M_{N}$ for nonzero $V_{\mu N}$ and $\mu_{\nu_\mu}$ with $M_{Z'}$ =0.1 (top-left), 1 (top-right), 10 (bottom-left) GeV.} 
\end{figure}

We now focus on the heavy neutrino $N$ that couples to only one of the SM lepton doublets, $L_{\mu}$. This model depends on 7 parameters: the mass of heavy neutrino $M_{N}$, the mass of $Z'$ boson $M_{Z'}$, the muon 
active-to-heavy-neutrino transitional magnetic moment $\mu_{\nu_{\mu}}$, heavy neutrino mixing angle $V_{\mu N}$ and the $Z'$ coupling with quarks $g_q$, with charged leptons $g_l$ and with neutrinos $g_\nu$. The rest of parameters are set at zero.

The branching ratios for the BM-I with three different $Z'$ masses are shown in Fig.~\ref{fig6}.
The branching ratios vary with $M_N$ and once the mass $M_N$ is above a
particular threshold, the corresponding new channel appears, 
e.g., $N\to \nu_\mu \tau^+ \tau^-$ appears when $M_N > 2 m_\tau$.
For $M_{Z'}<M_{N}$ the two-body decay of $N\rightarrow \nu _{\alpha} Z'$  
dominates, followed by the $Z'$ boson decay into 
SM fermions $Z'\rightarrow f \bar f$ 
($f = \nu, l, u, d$), according to the mass of $Z'$. 
The branching ratios of the heavy neutrino $N$ through such a 2-body decay 
can be converted to the 3-body one using 
\[
{\rm BR}(N \xrightarrow{Z'} \nu~X~X )=
{\rm BR} (N\rightarrow \nu Z') \times {\rm BR} (Z'\rightarrow X X).
\]

\begin{figure}[th!]
	\centering
	\includegraphics[width=16cm,height=8cm]{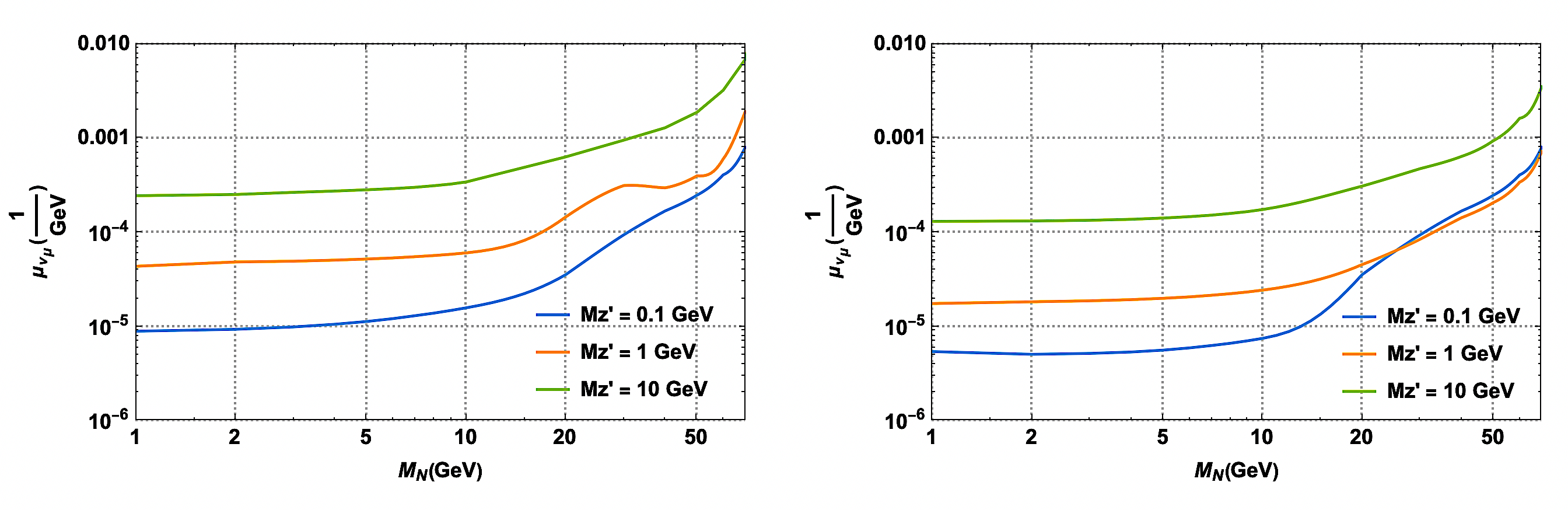}
	\caption{ \small \label{fig7}
    Sensitivity reach to $\nu_\mu$-to-heavy-neutrino transitional magnetic moment $\mu_{\nu_\mu}$ at FASER$\nu$ as a function of the heavy neutrino mass $M_N$ with different $Z'$ mass.
    Right: considering quarks and charged leptons in the the final state;   left: considering only charged leptons in the final state.}
\end{figure}

The sensitivity curves for the BM-I are shown in Fig.~\ref{fig7}. The right panel considers that the detector could probe both hadron and 
charged-lepton final states, while the left panel only considers the charged-lepton final state. By comparing the two plots we can see 
that the right panel shows better sensitivity. In both panels, the sensitivity reach on $\mu_{\nu_\mu}$ is the best at very small 
$M_{Z'}$ and $M_{N}$ region. For $M_{Z'} = 0.1 \rm~GeV $ and  $M_{N} = 1 \rm~GeV $ the sensitivity reach of $\mu_{\nu_\mu}$ is about 
$\sim 5\times10^{-6} \, {\rm GeV}^{-1}$ 
(see the right panel of Fig~\ref{fig7}), and 
$\sim 10^{-5} \, {\rm GeV}^{-1} $ (left panel of Fig.~\ref{fig7}). 
The left panel of Fig.~\ref{fig7} depends on 
$N\rightarrow \nu_\mu l^+ l^-$, from which we can see that
the change in the curve pattern can be read from the branching ratio curves in Fig.~\ref{fig6}. 
For $M_{Z'}=$ 0.1 GeV, $N\rightarrow\nu_\mu \nu \bar \nu$ is the dominant decay channel, while 
for other 2 cases $M_{Z'} = 1$ and 10 GeV $N \rightarrow \nu_\mu +
\rm hadrons$ becomes dominant.
The higher $M_{Z'}$ and $M_N$, the weaker the limit on $\mu_{\nu_\mu}$
will be.
Also, among all neutrino flavors, $\mu_{\nu_\mu}$ has the best 
sensitivity compared to other two benchmark models.

\subsection{Benchmark Model-II (BM-II): 
Heavy Neutrino $N$ and $Z'$  mixed with $e$ and $\nu_e$}

In this BM-II, the heavy neutrino coupled with the $L_e$ doublet. Similar to BM-I, this BM-II also depends on 7 parameters. Here the 
heavy neutrino has the non-zero mixing angle $V_{e N}$ with the SM leptons, and has the non-zero active-to-heavy neutrino transitional magnetic moment $\mu_{\nu_e}$.

\begin{figure}[th!]
	\centering
	\includegraphics[width=16cm,height=14cm]{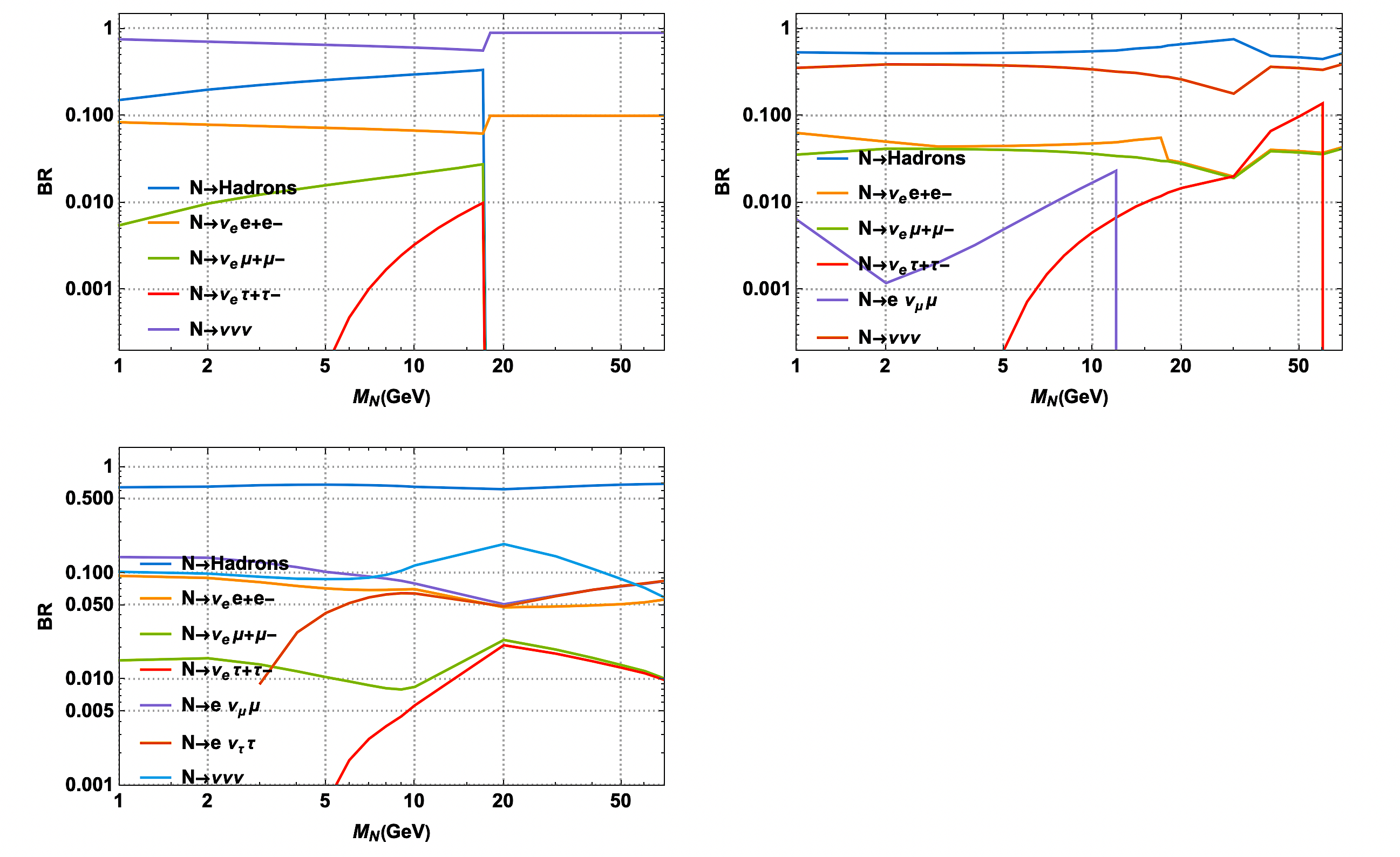}
	\caption{ \small \label{fig8}
    Dominant branching fractions of the heavy neutrino $N$ as a function of its mass $M_{N}$ for nonzero $V_{e N}$ and $\mu_{\nu_e}$  
    with $M_{Z'}$ =0.1 (top-left), 1 (top-right), 10 (bottom-left) GeV.}
\end{figure}
The branching ratios of $N\rightarrow \nu_e~X~X$ are shown in Fig.~\ref{fig8} with three different $M_{Z'}$. When the mass of $M_{N}>M_{Z'}$, the two-body decay of heavy neutrino 
$N\rightarrow \nu Z'$ dominates, followed by the $Z' \to f \bar f $.
\begin{figure}[h!]
	\centering
	\includegraphics[width=16cm,height=8cm]{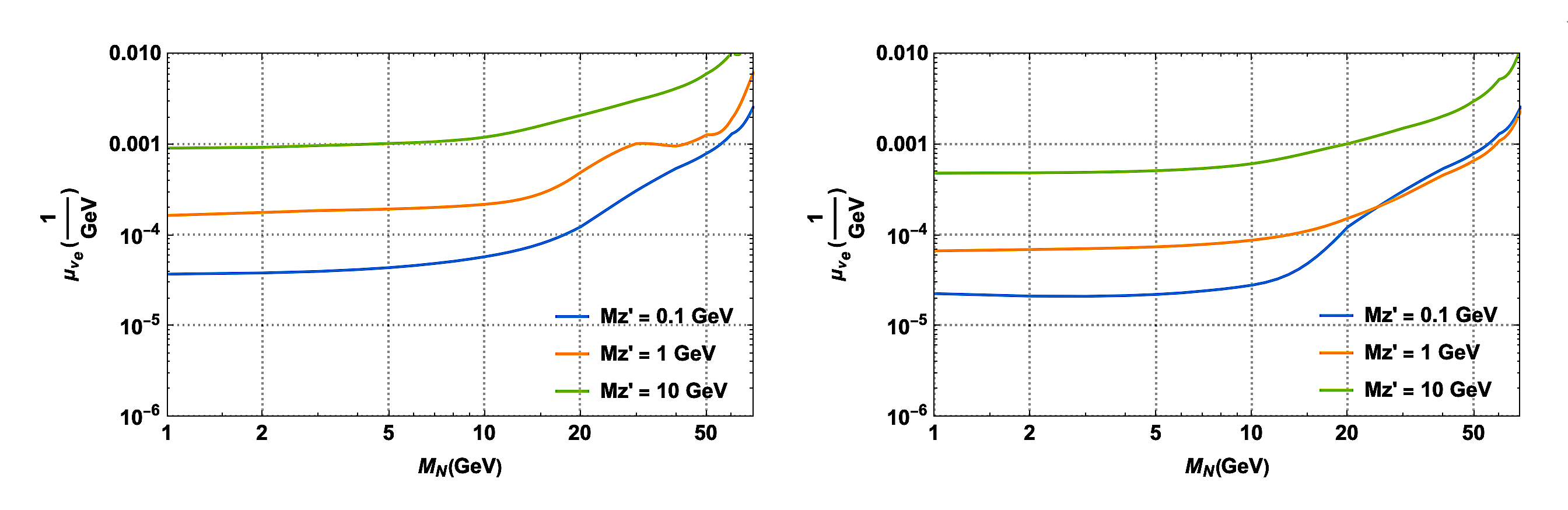}
	\caption{ \small \label{fig9}
Sensitivity reach to the $\nu_e$-to-heavy-neutrino transitional magnetic moment $\mu_{\nu_e}$ at FASER$\nu$ as a function of the heavy neutrino mass $M_N$ with different $Z'$ mass.
Right: considering both quarks and charged leptons in the final state; left: considering only charged leptons in the final state.}
\end{figure}

The sensitivity reach on  $\mu_{\nu_{e}}$ versus $M_{N}$ with different $M_{Z'}$ is shown in Fig.~\ref{fig9}. The sensitivity reach on $\mu_{\nu_{e}}$ is the best at very small $M_{N}$ and $M_{Z'}$ and is 
around $\sim 2\times10^{-5} \, {\rm GeV}^{-1} $ (right panel of Fig.~\ref{fig9}) and 
$\sim 5\times10^{-5}\, {\rm GeV}^{-1} $ (left panel of Fig.~\ref{fig9}). The right and left panels of Fig.~\ref{fig9} follow the same notation as Fig.~\ref{fig7} of BM-I. The $\mu_{\nu_\mu}$ sensitivity curves are better than those of $\mu_{\nu_e}$ curves.

\subsection{Benchmark Model-III (BM-III): 
Heavy Neutrino $N$ and $Z'$ mixed with $\tau$ and $\nu_\tau$}
\begin{figure}[h!]
	\centering
	\includegraphics[width=16cm,height=14cm]{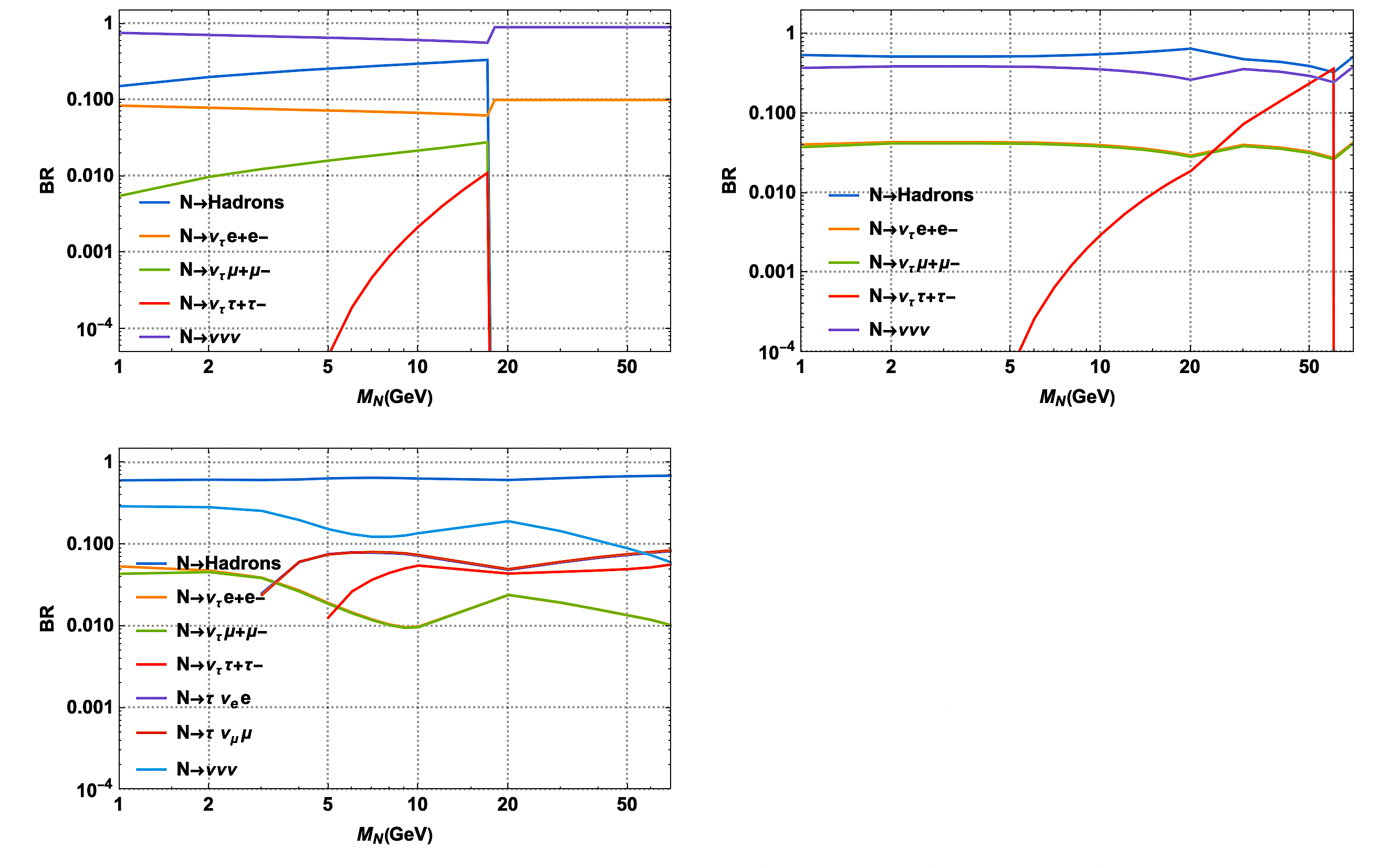}
	\caption{ \small \label{fig10}
Dominant branching fractions of the heavy neutrino $N$ as a function of its mass $M_{N}$ for nonzero $V_{\tau N}$ and $\mu_{\nu_\tau}$  with $M_{Z'}$ = 
0.1 (top-left), 1 (top-right), 10 (bottom-left) GeV.}
\end{figure}

In this model heavy neutrino $N$ only mixes with the $\tau$ lepton doublet through a nonzero mixing angle $V_{\tau N}$ and a non-zero transitional magnetic moment $\mu_{\nu_\tau}$. The branching ratios of $N\rightarrow \nu_\tau~X~X$ are shown in Fig.~\ref{fig10}. For $M_{Z'}=$ 0.1 GeV, $N\rightarrow\nu_\tau \nu \bar \nu$ is the dominant decay channel, while 
for other 2 cases $M_{Z'} = 1$ and 10 GeV $N \rightarrow \nu_\tau +
\rm hadrons$ becomes dominant.
We can see similar pattern for the heavy neutrino decay in the BM-I 
and BM-II.

\begin{figure}[h!]
	\centering
	\includegraphics[width=16cm,height=8cm]{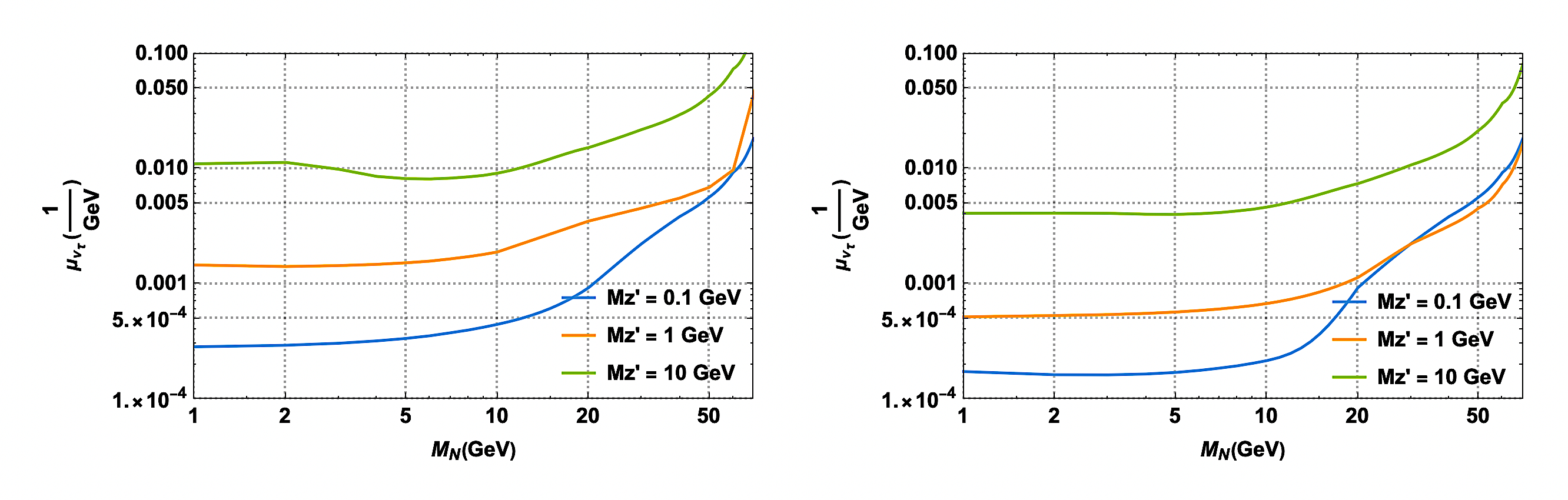}
	\caption{ \small \label{fig11}
Sensitivity reach to the $\nu_\tau$-to-heavy-neutrino transitional magnetic moment $\mu_{\nu_\tau}$ at FASER$\nu$ as a function of the heavy neutrino mass $M_N$ with different $Z'$ mass.
Right: considering quarks and charged leptons in the final state; left: considering only charged leptons in the final state.}
\end{figure}

The sensitivity reach for the active-to-heavy-neutrino transitional 
magnetic moment $\mu_{\nu_\tau}$ is depicted in Fig.~\ref{fig11}.
Among all three benchmark models, BM-III has the least sensitivity 
towards the transitional magnetic moment,
simply because of the small incoming tau-neutrino flux.
The left and right panels of Fig.~\ref{fig11} follow the same convention
as those of BM-I and BM-II. The sensitivity reach on $\mu_{\nu_\tau}$ 
is the best at very small $M_{N}$ and $M_{Z'}$ and is around $\sim 2\times10^{-4} \, {\rm GeV}^{-1}$ (right panel of figure \ref{fig11}) and $\sim 3\times10^{-4} \, {\rm GeV}^{-1}$ (left panel of figure \ref{fig11}). 
Among all three benchmark models, BM-I has the best sensitivity reach compared to the other two.  It is due to the difference in the flux of 
each neutrino flavor.

\section{Conclusions}

In this paper, we have studied the active-to-heavy-neutrino transitional magnetic moment which arises from the $Z'$ boson interactions between
the heavy and active neutrinos at FASER$\nu$. 
We consider the neutrino-nucleon scattering as the production process of the heavy neutrino, and we see that if the heavy neutrino can decay 
into charged leptons or hadrons, it can be identified with negligible
backgrounds inside the FASER$\nu$ detector. 
We investigated the advantage of FASER$\nu$ in a wide mass range 
search for the heavy neutrino $N$ and $Z'$ to determine the flavor dependence of the coupling between the active neutrinos and the 
heavy neutrino, for which we found that FASER$\nu$ is sensitive to $\mu_{\nu_\alpha}$ because of large incoming neutrino flux. 

In this study we have considered three benchmark models in which 
the heavy neutrino couples to $L_\mu$, $L_e$ or $L_\tau$ SM doublet. 
Among all the three models, the BM-I in which the heavy neutrino couples 
to $L_\mu$ has the best sensitivity reach compared to the other two.
We obtained sensitivity curves by considering two different final states: (i) charged leptons and (ii) charged leptons and hadrons. 
The second final state provides better sensitivity reach on $\mu_{\nu_\alpha}$. 
While the FASER$\nu$ can achieve the best sensitivity at small 
$M_{Z'}$ and $M_{N}$ regime, the sensitivity limits get weaker 
with the increment of heavy neutrino and $Z'$ masses. We did not include antineutrinos in our study; if we had included antineutrinos, the sensitivity curves could be improved by a factor of 2. 

A study of the magnetic dipole interactions between the active neutrinos and new sterile states at emulsion and liquid argon experiments 
was reported in \cite{Ismail:2021dyp}, which considered 
the up-scattering of neutrinos on electrons produces an electron-recoil signature that can probe new regions of parameter space at the
High Luminosity LHC (HL-LHC), in particular for liquid argon detectors due to low momentum thresholds. They also considered the decay of the sterile neutrino through the dipole operator, which leads to a photon that
could be displaced from the production vertex. 
Bohr magneton $\mu_B$ is 
given by $\mu_B = e \hbar /2m_e$, which is 
about $3 \times 10^{2}\;{\rm GeV}^{-1}$ in natural units. 
Reference~\cite{Ismail:2021dyp} obtained the 90\% C.L. contour curve for the active-to-heavy-neutrino transitional magnetic moment with heavy neutrino mass $M_N$ ranging from $10^{-3}~\rm GeV-10~ GeV$, the 
Forward Physics Facility (FPF) detectors can reach down to dipole coupling strengths of 
a few $10^{-9}\mu_B$ ($\sim 10^{-6}\rm~GeV^{-1}$) for $\mu_{\nu_e}$, 
$\sim10^{-9}\mu_B$(a few $10^{-7}\rm~GeV^{-1}$) for $\mu_{\nu_\mu}$,
and a few $10^{-8}\mu_B$($10^{-5}\rm~GeV^{-1}$) for $\mu_{\nu_\tau}$. According to \cite{Ismail:2021dyp}, starting at $M_N\sim$ $10^{-1}$ GeV, the sensitivity weakens.

\section*{Acknowledgements}
The work was supported in parts by the MoST of Taiwan with the grant no.110-110-2112-M-007-017-MY3.
 
\appendix
\section{Angular distributions of the heavy neutrino}
\label{app}
\begin{figure}[th!]
	\centering
	\includegraphics[width=16cm,height=14cm]{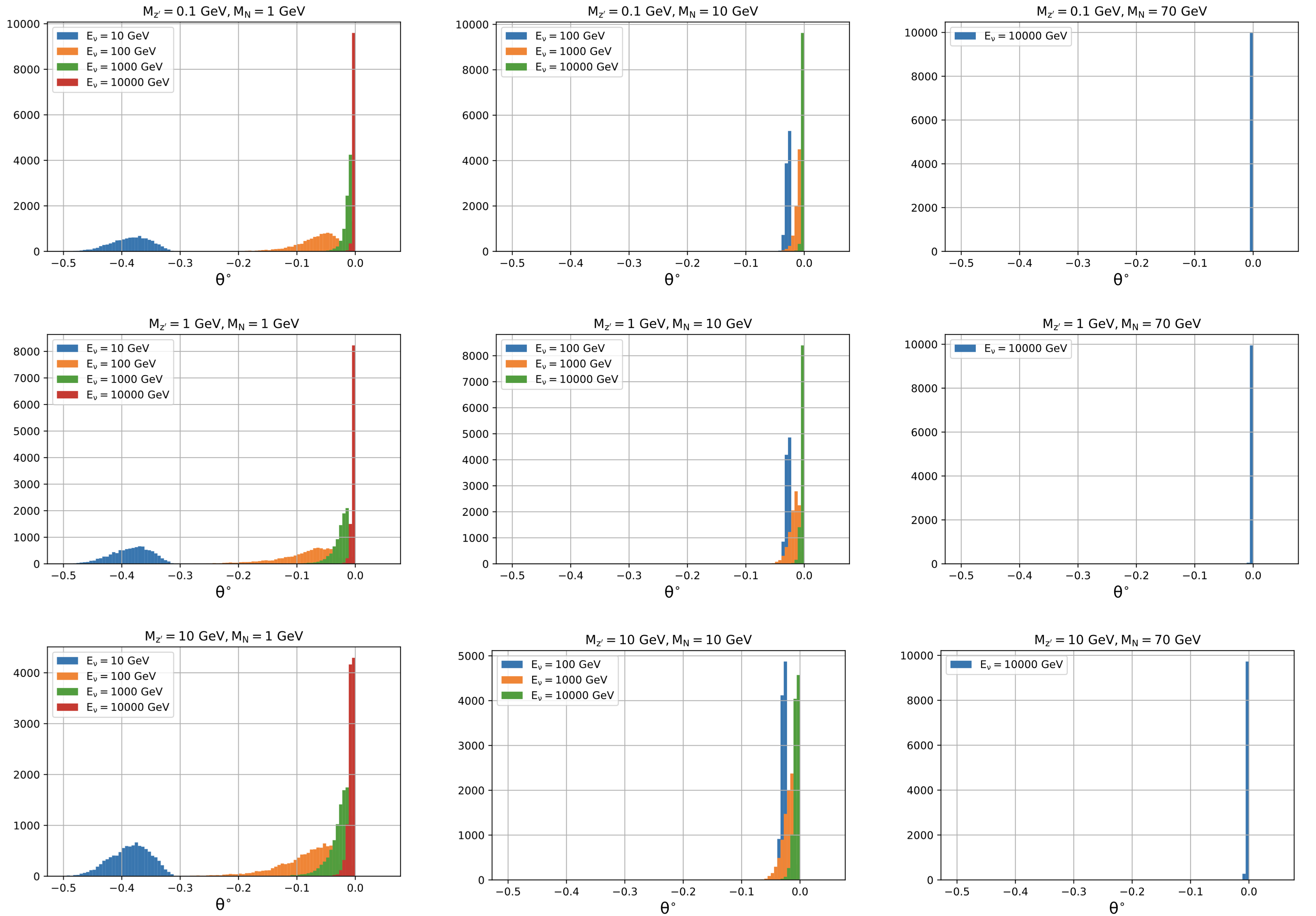}
	\caption{ \small \label{fig12}
Angular distributions of the heavy neutrino $N$ with different 
$M_{Z'}$ and different neutrino energy $E_{\nu}$.}
\end{figure}

Angular distribution curves are obtained by using the $\rm MadGraph5aMC@NLO$ \cite{Alwall:2011uj,Alwall:2014hca,Alwall:2014bza} 
and extracting the information from the output $\rm ``.lhe"$ file.
The angular distribution of the heavy neutral neutrino  
produced at FASER$\nu$ through the neutrino-nucleon scattering $\nu A\rightarrow N A$ is shown in Fig.~\ref{fig12}. 
Here we show the distributions with three $Z'$ masses $M_{Z'}$ = 0.1 ,1 10 GeV and with different sets of neutrino energies from 
10 GeV to 10 TeV. The distribution curves are approaching to 
`zero' by the increment of incoming active neutrino energy $E_{\nu}$, 
and the heavy neutrino Mass $M_{N}$.

\end{document}